\documentclass[prl, showpacs, twocolumn, superscriptaddress]{revtex4}
\usepackage{graphics}

\newcommand{\bra}{\left\langle}
\newcommand{\ket}{\right\rangle}

\newcommand{\pder}[2]{\frac{\partial #1}{\partial  #2}}

\newcommand{\der}[2]{\frac{\mathrm{d} #1}{\mathrm{d}  #2}}

\newcommand{\pc}{f_{\rm c}}

\newcommand{\ep}{\epsilon}
\newcommand{\fp}{f^{\rm p}}

\newcommand{\tmicro}{\tau_{\rm m}}
\newcommand{\tmacro}{\tau_{\rm M}}

\newcommand{\tp}{\tau_{\rm p}}

\newcommand{\e}{{\rm e}}

\newcommand{\vecr}{{\mathbf r}}
\newcommand{\vecp}{{\mathbf p}}
\newcommand{\vecE}{{\mathbf E}}
\newcommand{\vecR}{{\mathbf R}}
\newcommand{\vecP}{{\mathbf P}}
\newcommand{\vecF}{{\mathbf F}}
\newcommand{\vece}{{\mathbf e}}
\newcommand{\vecV}{{\mathbf V}}
\newcommand{\vecrb}{{\mathbf u}}
\newcommand{\vecpb}{{\mathbf g}}

\newcommand{\mb}{ m^{\rm b}}

\begin{document}

\title{Microscopic description of the equality between 
violation of fluctuation-dissipation relation and energy 
dissipation}

\author{Hiroshi Teramoto}
\email[Electronic address: ]{teramoto@mns2.c.u-tokyo.ac.jp}
\affiliation
{Department of Pure and Applied Sciences, University of Tokyo, Komaba, Tokyo 153-8902, Japan}

\author{Shin-ichi Sasa}
\email[Electronic address: ]{sasa@jiro.c.u-tokyo.ac.jp}
\affiliation{Department of Pure and Applied Sciences, University of Tokyo, Komaba, Tokyo 153-8902, Japan}

\date{\today}

\begin{abstract}
In systems far from equilibrium, the fluctuation-dissipation relation 
is violated due to the lack of detailed balance. Recently, 
for a class of Langevin equations, it has been proved that this 
violation is related to  energy dissipation as an equality
[T. Harada and S. Sasa, Phys. Rev. Lett., in press; cond-mat/0502505]. 
We  provide a microscopic description of this equality
by studying  a non-equilibrium colloidal system on the basis 
of classical mechanics with some physical assumptions.
\end{abstract}

\pacs{05.40.-a, 05.20.Gg}

\maketitle

%%%%%%% Introduction %%%%%%%%%%%%%%%%

%%% introduction

% general 

The construction of non-equilibrium statistical mechanics that is useful 
for systems far from equilibrium is a fundamental problem in theoretical
physics. Let us recall that linear response theory had been formulated 
for  the microscopic description of universal relations established by 
Einstein, Nyquist, and Onsager \cite{lrt}. 
Thus, it is a significant first step to provide  a
microscopic description of  formulae  derived 
phenomenologically for systems far from equilibrium. 

% Harada-Sasa

Recently, an interesting equality relating  the violation of 
the fluctuation-dissipation relation (FDR) with  energy dissipation 
has been found for Langevin equations under non-equilibrium conditions
\cite{Harada,Harada2}. Here, note that the FDR is a fundamental 
relation proved in a linear response regime around equilibrium 
states \cite{lrt}, but it is 
violated in systems far from equilibrium \cite{fdr-vio}. Recent 
studies have revealed that the idea of effective temperature 
is useful to characterize the FDR violation for glassy systems 
\cite{g1,g2,g3} and steady state systems \cite{s1,s2,s3}. 
In contrast to these studies,  this  equality claims that the 
FDR violation is characterized by an energetic quantity.

% question 

The equality has been proved for a wide class of Langevin systems 
including many-body systems, time-dependent potential systems, and 
systems in contact with many heat reservoirs \cite{Harada2}.
We then search for such an equality in more general 
non-equilibrium systems that are not necessarily described 
by a Langevin equation. 
To this end, we start with investigating  a non-equilibrium 
colloidal system on the basis of classical mechanics. Because it is
highly expected that the motion of a colloidal particle is described 
by a Langevin equation,  the FDR violation should be related 
to the energy dissipation even if we describe the system on the
basis of classical 
mechanics. Here, the energy dissipation is given by the  energy 
transfer from the center of mass of the colloidal particle to the 
other mechanical degrees of freedom  in the classical mechanical
description.   In this paper, we derive an expression of the FDR 
violation for the classical mechanical description of  the system
far from equilibrium. By considering the physical conditions of the 
system, we rederive the equality reported in Ref. \cite{Harada} 
from the expression of the FDR violation.

%%%%%%%%%%%%%%%%%%  model %%%%%%%%%%%%%%%%%%%%%%%% 

\paragraph{Model:}

% physical situation

Specifically, we study a system of one colloidal particle suspended 
in a three-dimensional liquid confined in a region where $ -L/2 \le y \le L/2$ 
and $ -L/2 \le z \le L/2$. Periodic boundary conditions are imposed 
in the $x$ direction. Let $Z=(\vecR,\vecP)$ be the position and momentum 
of the center of mass of the colloidal particle. The colloidal particle 
is driven by an external field $\vecE(t)=(E_0+\ep \fp(t),0,0)$ and is 
subject to a spatially periodic potential $U_0(\vecR)$, where $E_0$ 
is a constant force to realize a non-equilibrium steady state and
$\ep \fp(t)$ is a small probe force to investigate the response of the 
system in the steady state. 

% mechanical description (variables)

The center of mass of the colloidal particle interacts with the  
other mechanical degrees of freedom of the colloid-liquid system.
Their dynamical degrees of freedom are 
represented by $Y=(\vecr_1,\vecp_1,\cdots,\vecr_N,\vecp_N)$.
Furthermore, we introduce  thermostated walls as a network of  
boundary particles, whose dynamical degrees of freedom are represented by 
$B=(\vecrb_1,\vecpb_1,\cdots, \vecrb_{N'}, \vecpb_{N'})$. These
boundary particles are localized around the boundaries given by 
$y=\pm L/2$ and $z=\pm L/2$, and interact with both the colloidal 
particle and the molecules in the liquid. 

% mechanical description (Hamiltonian and temperature)

All the interaction potentials including $U_0(\vecR)$ are 
represented by a Hamiltonian $H(Z,Y,B)$. Note that a
constant force $E_0$ cannot be described by the Hamiltonian
because $E_0$ is regarded as a non-potential force
due to the periodic boundary condition in the $x$ direction.
The result presented below does not depend on the   
details of the system. We assume that in the equilibrium case ($E_0=0$),
the statistical properties of a variable set $(Z,Y,B)$ are 
described by a canonical distribution with  temperature $T$ of 
the thermostated walls. We also assume that 
there exists a steady distribution when $E_0 \not =0$. 
In order to control the temperature of a finite system, we introduce a 
Nos\'e-Hoover thermostat only for the boundary particles \cite{Nose}. 
The thermostat may introduce  an unphysical effect, but we expect that 
the influence of the thermostat vanishes in the limit 
$L, N \to \infty$ with $N/L^3$ fixed. Furthermore, when the mathematical 
rigor is not critically taken into account, the analysis developed 
below can be applied to a Hamiltonian system without a thermostat 
in this limit. 

% mechanical description (equation of motion) 

In this mechanical system, the motion of all the particles 
is described by
\begin{eqnarray}
\der{\vecP}{t} &= & \vecE - \pder{H}{\vecR}, \\
\der{\vecp_i}{t} &= &  - \pder{H}{\vecr_i},  \\
\der{\vecpb_i}{t} &= &  - \pder{H}{\vecrb_i}-\lambda_i \vecpb_i, \\
\tau \der{\lambda_i}{t} &= &  \frac{\vecpb_i^2}{\mb_i}- 3T,
\end{eqnarray}
with $\vecP=M d\vecR/dt$, $\vecp_i=m_i d\vecr_i/dt$, 
and $\vecpb_i=\mb_i d \vecrb_i/dt$, where $M$, $m_i$, and $\mb_i$ are
the masses of the corresponding variables.  We express these equations as
\begin{equation}
\der{\Gamma}{t} = \Phi(\Gamma,t)+\ep\Phi_1(t),
\label{eq1}
\end{equation}
where $\Gamma=(Z,Y,B,\lambda)$ with $\lambda=(\lambda_1,\cdots,\lambda_{N'})$.

%%%%%%%%%%% analysis %%%%%%%%%%%%%%%

\paragraph{Formal analysis:}

We consider the time-dependent distribution function $f(\Gamma,t)$
with an initial condition that $f(\Gamma,t_0)=\pc(\Gamma)$, where
\begin{equation}
\pc(\Gamma)=\frac{1}{Z}\e^{-\beta H_0(\Gamma)
-\beta \tau \sum_{i=1}^{N'}\lambda_i^2/2}
\label{init-dis}
\end{equation}
with $H_0(\Gamma) \equiv H(\Gamma)-U_0(\vecR)$. 
$Z$ is the normalization constant. 
The distribution function $f(\Gamma,t)$ satisfies  the equation
\begin{equation}
\pder{f(\Gamma,t)}{t}
= -\pder{}{\Gamma} \left( \Phi(\Gamma) f(\Gamma,t) \right)
  -\ep \pder{}{\Gamma} \left( \Phi_1(t) f(\Gamma,t) \right).
\label{Lio}
\end{equation}
Note that $\pc(\Gamma)$ is the stationary solution of (\ref{Lio}) 
when $\vecE(t)=0$ and $U_0(\vecR)=0$. 

We first set
\begin{equation}
f(\Gamma,t)=\pc(\Gamma)\e^{\beta \phi(\Gamma,t)}.
\label{set}
\end{equation}
Substituting this expression into (\ref{Lio}), we obtain 
\begin{eqnarray}
\pder{\phi(\Gamma,t)}{t}
&=& -\Phi(\Gamma)\pder{\phi(\Gamma,t)}{\Gamma}
+\vecF(\vecR)\cdot\pder{H_0}{\vecP} \nonumber \\
& & 
+\ep \fp(t)\pder{H_0}{P_x} -\ep \fp(t)\pder{\phi}{P_x},
\label{phieq}
\end{eqnarray}
where $\vecF(\vecR)$ is defined as
\begin{equation}
\vecF(\vecR)\equiv E_0\vece_x-\pder{U_0(\vecR)}{\vecR}.
\end{equation}
We solve (\ref{phieq}) formally as
\begin{eqnarray}
\phi(\Gamma,t)
&=&
\int_{t_0}^t ds \e^{\Lambda(s-t)} 
\left[\vecF(\vecR)\cdot \pder{H_0}{\vecP} \right. \nonumber \\
& &
+\left. \ep \fp(s)\pder{H_0}{P_x}-\ep\fp(s)\pder{\phi}{P_x}
\right],
\label{phi1}
\end{eqnarray}
where we have defined
\begin{equation}
\Lambda \equiv \Phi(\Gamma) \pder{}{\Gamma}.
\end{equation}

% expression 

In the argument below, we fix $t$ and denote the solution
of (\ref{eq1}) with $\ep=0$, which satisfies
$\Gamma(t)=\Gamma$, as $\Gamma(s)$ with $t_0 \le s \le t$. 
That is, $\Gamma(s)$
is regarded as a function of $\Gamma$. 
Then, for an arbitrary function $A(\Gamma)$, we obtain
\begin{equation}
\der{A(\Gamma(s))}{s} = \Lambda A(\Gamma')|_{\Gamma'=\Gamma(s)}.
\end{equation}
Therefore, noting $A(\Gamma(t))= A(\Gamma)$, we write
\begin{equation}
\e^{\Lambda (s-t)}A(\Gamma)=A(\Gamma(s)),
\label{e-l}
\end{equation}
where the right-hand side is regarded as a function of $\Gamma$ according
to the rule mentioned above. Using this expression, we rewrite (\ref{phi1})
as 
\begin{eqnarray}
\phi(\Gamma,t)
&=&
\int_{t_0}^t ds 
\left[\vecF(\vecR(s))\cdot \vecV(s) \right. \nonumber \\
&+ &
\left. \ep \fp(s)\pder{H_0(\Gamma(s))}{P_x(s)}
-\ep\fp(s)\pder{\phi(\Gamma(s),t)}{P_x(s)}
\right],
\label{phi2}
\end{eqnarray}
where we have defined $\vecV(s)\equiv \vecP(s)/M$.

% expansion of phi

Now, let us expand $\phi(\Gamma,t)$ as
\begin{equation}
\phi(\Gamma,t)=\phi^{(0)}(\Gamma,t)+\ep\phi^{(1)}(\Gamma,t)+O(\ep^2).
\label{expa}
\end{equation}
Substituting this expression into (\ref{phi2}), we arrange the terms 
according to the powers of $\ep$. 
From the terms independent of $\ep$, we obtain 
\begin{equation}
\phi^{(0)}(\Gamma,t)= W(\Gamma,t_0,t;t)
\label{sol:0}
\end{equation}
with 
\begin{equation}
W(\Gamma,t_0,s;t)
=
\int_{t_0}^s d\vecR(s')\cdot \vecF(\vecR(s')),
\label{Wdef}
\end{equation}
where $W(\Gamma,t_0,s;t)$ represents the accumulated work done by $\vecF$ 
during the time interval $[t_0,s]$ for the trajectory satisfying 
$\Gamma(t)=\Gamma$. 
The expression of the distribution function (\ref{set}) 
with (\ref{sol:0}) and (\ref{Wdef}) is similar to that proposed 
by Zubarev \cite{Zubarev} and McLennan \cite{Maclennan}. 
Next, the terms proportional to  $\ep$ yield 
\begin{equation}
\phi^{(1)}(\Gamma,t)= \int_{t_0}^t ds  \fp(s) \left(
V_x(s)-\pder{W(\Gamma,t_0,s;t)}{P_x(s)}
\right).
\label{sol:1}
\end{equation}

% average 
Furthermore, the average of $A(\Gamma)$  by the distribution 
function $f(\Gamma,t)$,
\begin{equation}
\bra A(\Gamma(t)) \ket_{\ep,t_0} \equiv \int d\Gamma f(\Gamma,t)A(\Gamma),
\end{equation}
is expanded in $\ep$ as
\begin{equation}
\bra A(\Gamma(t)) \ket_{\ep,t_0} = \bra A(\Gamma(t)) \ket_{t_0}^{(0)}
+ \ep \bra A(\Gamma(t)) \ket_{t_0}^{(1)}+O(\ep^2).
\end{equation}
Then, from (\ref{set}), (\ref{expa}), and (\ref{sol:1}), we obtain
\begin{eqnarray}
\bra V_x(t)\ket_{t_0}^{(1)} 
&=& 
\beta \int_{t_0}^t ds  
\fp(s) \bra V_x(t) V_x(s)\ket_{t_0}^{(0)} \nonumber \\ 
&- & 
\beta \int_{t_0}^t ds  
\fp(s) \bra V_x(t)\pder{W(\Gamma(t),t_0,s;t)}{P_x(s)}
\ket_{t_0}^{(0)} .
\label{fvio}
\end{eqnarray}

%% remark 

Here, we remark on the steady state of the system. It can be expected 
that $\bra A(\Gamma(t))\ket_{\epsilon,t_0}$ becomes  an averaged value 
in the steady state when we take the limit $t_0 \to -\infty$. However, 
$f(\Gamma,t)$ itself is divergent in this limit, as seen from (\ref{sol:0}) 
and (\ref{Wdef}). In this paper, we  consider  the limit for the averaged 
quantities, but we do not study the singularity of the  distribution 
function itself.

Taking this into consideration, 
setting $t_0 \to -\infty$ in (\ref{fvio}), 
we obtain the formula
\begin{equation}
C(t-s)=TR(t-s) 
+\lim_{t_0 \to -\infty} 
\bra V_x(t) \pder{W(\Gamma(t), t_0,s;t)}{P_x(s)}\ket_{t_0}^{(0)}
\label{goal}
\end{equation}
for $t >s $. In this formula, 
the time correlation function $C(t-s)$ and the response function
$R(t-s)$ are defined as
\begin{eqnarray}
\bra V_x(t)\ket_{-\infty}^{(1)}&=& 
\int_{-\infty}^t ds R(t-s) \fp(s), \\
C(t-s) &=&  \bra V_x(t) V_x(s) \ket_{-\infty}^{(0)}. 
\end{eqnarray}
We also define $R(t)=0$ for $t <0$ from the causality.

To this point,  no approximation is involved.
Then, in the equilibrium case ($E_0=0$),  we derive the  FDR
\begin{equation}
C(t-s)=T R(t-s)
\label{FDT}
\end{equation}
for $t >s $. This derivation is straightforward when we note that
$W(\Gamma(t), t_0,s;t)=-U_0(\vecR(s))+U_0(\vecR(t_0))$
in this case. Here, we have used 
\begin{equation}
\lim_{t_0 \to  - \infty}\bra V_x(t) \pder{U_0(\vecR(t_0))}{P_x(s)}
\ket_{t_0}^{(0)}=0.
\end{equation}
On the other hand, when $E_0 \not= 0$, the second term on
the right-hand side of (\ref{goal}) takes a finite (nonzero) value  
in general. 
Thus, (\ref{goal}) provides an expression of the FDR  violation 
in the mechanical description.

%%%%%%%%%%%%%%%%%%%%%%%%%%%

\paragraph{Physical consideration:}

Let us estimate the quantity  
$\partial W(\Gamma(t),t_0,s;t)/\partial P_x(s)$ in (\ref{goal}) 
for the system considered in this study. Our estimation is based on
the two important assumptions on the time scales. 
First, let $\tmicro$ 
be the slowest time scale of phenomena that  a set of 
variables $(Y,B,\lambda)$ exhibits. This time scale $\tmicro$ is 
expected to be much smaller than the relaxation time of the 
velocity of the colloidal particle, which is denoted as $\tmacro$. 
Considering this physical expectation, we assume that there exists a
time scale $\Delta_1$ satisfying
\begin{equation}
\tmicro \ll \Delta_1 \ll \tmacro.
\label{first}
\end{equation}
We choose such a time scale $\Delta_1$ and hereinafter fix it.
Using this $\Delta_1$, we define the time-averaged quantity as
\begin{equation}
\overline{A(\Gamma(s))} \equiv \frac{1}{\Delta_1}
\int_{s-\Delta_1/2}^{s+\Delta_1/2} ds'A(\Gamma(s')).
\end{equation}

The second assumption on the time scale of the system is that the 
relaxation time of the velocity $\tmacro$ is much smaller than 
the typical time scale of the position variation $\tp$ that is
determined by the  characteristic length of the potential $U_0(\vecR)$. 
That is, we can choose a time scale $\Delta_2$ satisfying
\begin{equation}
\tmacro \ll \Delta_2 \ll \tp.
\label{second}
\end{equation}

Now, we apply the perturbation $\delta V_x(s)$ to the phase space
point $\Gamma(s)$. As a result of the perturbation, the trajectory
$\Gamma(s')$ with $ s' \le s $ changes to $\Gamma(s')+\delta \Gamma(s')$,
and this change yields the additional work $\delta W(\Gamma(t),t_0,s;t)$
from (\ref{Wdef}). It seems reasonable to assume that the relaxation 
time of the work rate $\vecV(s')\cdot \vecF(\vecR(s')) $ is of the  order 
of $\tmacro$. Thus, from the second assumption (\ref{second}), we can estimate
\begin{equation}
\delta W(\Gamma(t),t_0,s;t)
\simeq 
\int_{s-\Delta_2}^s ds' 
\delta (\vecV(s')\cdot \vecF(\vecR(s'))).
\label{dW}
\end{equation}
Here, assuming  $\overline{\vecV\cdot \vecF(\Gamma(s'))}
\simeq \overline{\vecV(s')}\cdot \overline{\vecF(\vecR(s'))}$,
we write
\begin{equation}
\vecV(s')\cdot \vecF(\vecR(s'))=
\overline{\vecV(s')}\cdot \overline{\vecF(\vecR(s')}+\eta(s'),
\label{etadef}
\end{equation}
where $\eta(s)$ represents the {\it fast part} of the work rate fluctuation
whose time scale is much smaller than $\Delta_1$. 
Using this, we express
\begin{equation}
\delta W(\Gamma,t_0,s;t)
\simeq 
\int_{s-\Delta_2}^s ds' 
\delta[\overline{\vecV(s')}\cdot \overline{\vecF(\vecR(s'))}+\eta(s')].
\label{dW2}
\end{equation}
From the second  assumption (\ref{second}), 
$\overline{\vecF(\vecR(s'))}$ in the integrand can be replaced with 
$\overline{\vecF(\vecR(s))}$, and the estimation 
\begin{equation}
\delta \overline{\vecV(s')}
=(\e^{-\frac{(s-s')}{\tmacro}}\delta V_x(s),0)
\end{equation}
seems reasonable in the time interval $[s-\Delta_2, s]$. 
Thus, the first term in (\ref{dW2}) can be rewritten as 
\begin{equation}
\overline{\vecF(\vecR(s))}
\int_{s-\Delta_2}^s ds' 
\delta \overline{\vecV(s')} 
\simeq 
\overline{F_x(\vecR(s))}
\delta V_x(s) \tmacro.
\label{dW3}
\end{equation}
From this result, we obtain the following  expression: 
\begin{equation}
\pder{W(\Gamma,t_0,s;t)}{P_x(s)}
\simeq \frac{\overline{F_x(\vecR(s))}\tmacro}{M}+
\int_{s-\Delta_2}^s ds' 
\pder{\eta(s')}{P_x(s)}.
\label{assump}
\end{equation}
Recalling the definition of $\eta$ in (\ref{etadef}) and the first 
assumption (\ref{first}), we expect that the second 
term of (\ref{assump}) has negligible correlation 
with $V_x(t)$. Further, when we assume that a friction force for  the 
the colloidal particle is given by $-\gamma \vecV$, $\gamma$ is estimated
as $M/\tmacro$. Using  these, the substitution of (\ref{assump}) into
 (\ref{goal}) leads to 
\begin{equation}
C(t-s)=T R(t-s)+ \frac{1}{\gamma}\bra V_x(t)F_x(s) \ket_{-\infty}^{(0)}
\label{FDT-vio}
\end{equation}
for $t >s$.

Let us define  the Fourier transform of $C(t-s)$ as
\begin{equation}
\tilde C(\omega)\equiv \int_{-\infty}^\infty dt' \e^{i \omega t'} C(t').
\end{equation}
Similarly, we define $\tilde R(\omega)$, while $\tilde R'(\omega)$ denotes
the real part of $\tilde R(\omega)$. Then, from (\ref{FDT-vio}), 
we obtain 
\begin{equation}
\int_{-\infty}^\infty \frac{d\omega}{2\pi}
[\tilde C(\omega)-2T \tilde R'(\omega)]
=\frac{1}{\gamma} J,
\label{Harada-Sasa}
\end{equation}
where $J=\bra  V_x F_x\ket_{-\infty}^{(0)}$ 
is the energy transfer rate from the center of mass of the colloidal
particle to the other degrees of freedom. This energy transfer
rate is interpreted as the energy dissipation ratio in the description 
of  the center of mass of the colloidal particle.
Thus, the equality given in (\ref{Harada-Sasa}) relates the FDR violation
with  energy dissipation, as presented for a Langevin equation
in Ref. \cite{Harada}. In this manner, we have rederived the equality
on the basis of the classical mechanical description.

%%%% discussion %%%%

\paragraph{Discussion:}

It might be possible to develop a theory for formalizing the 
above mentioned physical consideration. In such a theory, the 
equality given in (\ref{Harada-Sasa}) might be derived systematically 
from microscopic dynamics by  a calculation technique using
the separation of time scales. The construction of the theory 
is a future research subject.

Related to this subject, we remark that there exist many 
different expressions of the distribution function. 
For example, when we assume an  initial condition  involving 
$H$ instead of $H_0$ in (\ref{init-dis}), we obtain
a different expression of the distribution function. However,
in this case, we find that it is difficult to relate  the 
obtained expression of the FDR violation with the result 
for the Langevin equation. Note that the steady  distributions
for  both the initial conditions should be identical 
when the limit $t_0 \to -\infty$ is considered. This implies 
that there is an expression 
that can be treated in a simple manner. The clarification of  this 
might provide a key step in the systematic derivation 
of the  equality given in (\ref{Harada-Sasa}).

A more important question is whether  FDR violation can be expressed 
by a form useful for systems in which  a Langevin description is 
not effective. It should be noted  that one can derive an expression 
of FDR violation for any mechanical system in a similar manner to that 
used for deriving (\ref{goal}). 
The examples of  mechanical systems include electric conduction systems, 
sheared systems, and heat conduction systems. 
However, as mentioned above, the obtained expression  might have no 
direct relation with measurable energetic quantities. 
Thus, it is necessary to find a condition under which the FDR 
violation takes a physically useful form.

In conclusion, we provide a microscopic description 
for the equality given in (\ref{Harada-Sasa}) by analyzing the Hamiltonian
equation with the Nos\'e-Hoover thermostat 
at the boundaries. By examining questions arising from this study,
we wish to obtain a deep understanding of non-equilibrium
systems.

\begin{acknowledgments}
The authors thank K. Hayashi, T. Harada, and H. Tasaki for useful 
discussions. 
This work was supported by a grant from the Ministry of Education, 
Science, Sports and Culture of Japan (No. 16540337) and a grant 
from Research Fellowships of the Japan Society for the Promotion 
of Science for Young Scientists (No. 1711363).  
\end{acknowledgments}

%%%%%%%%%%%%%%%%%%%%
%%%% references %%%%
%%%%%%%%%%%%%%%%%%%%

\end{document}